\documentclass{optica-article}

\journal{opticajournal}

\articletype{Research Article}

\usepackage{booktabs,multirow,lineno,makecell}
\linenumbers

\begin{document}
\nolinenumbers
\title{Accelerated deep self-supervised ptycho-laminography for three-dimensional nanoscale imaging of integrated circuits}

\author{Iksung Kang,\authormark{1,8} Yi Jiang,\authormark{2} Mirko Holler,\authormark{3} Manuel Guizar-Sicairos,\authormark{3,4} A. F. J. Levi,\authormark{5} Jeffrey Klug,\authormark{2} Stefan Vogt,\authormark{2,9} and George Barbastathis\authormark{6,7,10}}

\address{\authormark{1}Department of Electrical Engineering and Computer Science, Massachusetts Institute of Technology, Cambridge, MA 02139, USA\\
\authormark{2}Argonne National Laboratory, Lemont, IL 60439, USA\\
\authormark{3}Paul Scherrer Institut, Forschungsstrasse 111, 5232 Villigen PSI, Switzerland\\
\authormark{4}Institute of Physics (IPHYS), Ecole Polytechnique Fédérale de Lausanne, Rte Cantonale, Lausanne 1015, Switzerland\\
\authormark{5}Department of Electrical and Computer Engineering, University of Southern California, Los Angeles, CA 90007, USA\\
\authormark{6}Department of Mechanical Engineering, Massachusetts Institute of Technology, Cambridge, MA 02139, USA\\
\authormark{7}Singapore-MIT Alliance for Research and Technology (SMART) Centre, 1 Create Way, Singapore 117543, Singapore\\
\authormark{8}Present address: Department of Molecular and Cell Biology, University of California, Berkeley, California 94720, USA\\
\authormark{9}\email{svogt@anl.gov}
\authormark{10}\email{gbarb@mit.edu}}




\begin{abstract}
Three-dimensional inspection of nanostructures such as integrated circuits is important for security and reliability assurance. Two scanning operations are required: ptychographic to recover the complex transmissivity of the specimen; and rotation of the specimen to acquire multiple projections covering the 3D spatial frequency domain. Two types of rotational scanning are possible: tomographic and laminographic. For flat, extended samples, for which the full 180 degree coverage is not possible, the latter is preferable because it provides better coverage of the 3D spatial frequency domain compared to limited-angle tomography. It is also because the amount of attenuation through the sample is approximately the same for all projections. However, both techniques are time consuming because of extensive acquisition and computation time. Here, we demonstrate the acceleration of ptycho-laminographic reconstruction of integrated circuits with $16$-times fewer angular samples and $4.67$-times faster computation by using a physics-regularized deep self-supervised learning architecture. We check the fidelity of our reconstruction against a densely sampled reconstruction that uses full scanning and no learning. As already reported elsewhere [Zhou and Horstmeyer, \textit{Opt. Express}, 28(9), pp. 12872-12896], we observe improvement of reconstruction quality even over the densely sampled reconstruction, due to the ability of the self-supervised learning kernel to fill the missing cone.
\end{abstract}

\section{Introduction}
Hard X-rays offer non-destructive visualization and metrology of nanoscopic details inside complex structures, such as Integrated Circuits (ICs). Short wavelength and long penetration depth of X-rays enable probing into the volumetric interiors of ICs. X-ray imaging instruments often incorporate the object rotation with respect to the X-ray illumination to improve the depth resolution of reconstructions. It is desirable to select the rotation geometry by carefully considering the objects' geometrical properties. For instance, for flat extended nanostructures like ICs, the object’s rotation axis can be oblique to the direction of synchrotron X-rays, \textit{i.e.} laminographic imaging~\cite{helfen2005high,helfen2011implementation}. For flat, extended samples, the oblique geometry keeps the amount of X-ray absorption and scattering by the structures approximately the same regardless of the object rotation, so that the volumetric interiors of ICs can be more reliably reconstructed. This makes a clear distinction from existing tomographic imaging methods~\cite{levine1999tomographic, tkachuk2006high,holler2014x,holler2017high}, where the strengths of absorption would vary across different rotation angles.

On the other hand, translational scanning of the object enables a larger field of view reconstruction. Originally proposed for Scanning Transmission Electron Microscopy (STEM), ptychography leverages lateral movement of either the object or the illumination to acquire several diffraction patterns from different lateral locations in order to add robustness to phase retrieval~\cite{hoppe1969beugung,hegerl1970dynamische,rodenburg1992theory,guizar2021ptychography} and to reconstruct a larger field-of-view object~\cite{rodenburg2004phase,rodenburg2007transmission,faulkner2004movable}. Objects are computationally retrieved from the ptychographic measurements by some well-established algorithms, such as ptychographic iterative engine (PIE)~\cite{rodenburg2004phase}, difference map (DM)~\cite{thibault2009probe}, least-square maximum likelihood (LSQ-ML)~\cite{odstrvcil2018iterative}, etc. Alternatively, ptychography may also be conducted in the Fourier domain by replacing the illumination with a set of plane waves incident at different angles~\cite{konda2020fourier,zheng2013wide,tian2014multiplexed}.

\begin{figure}[htbp!]
    \centering
    \includegraphics[width=\textwidth]{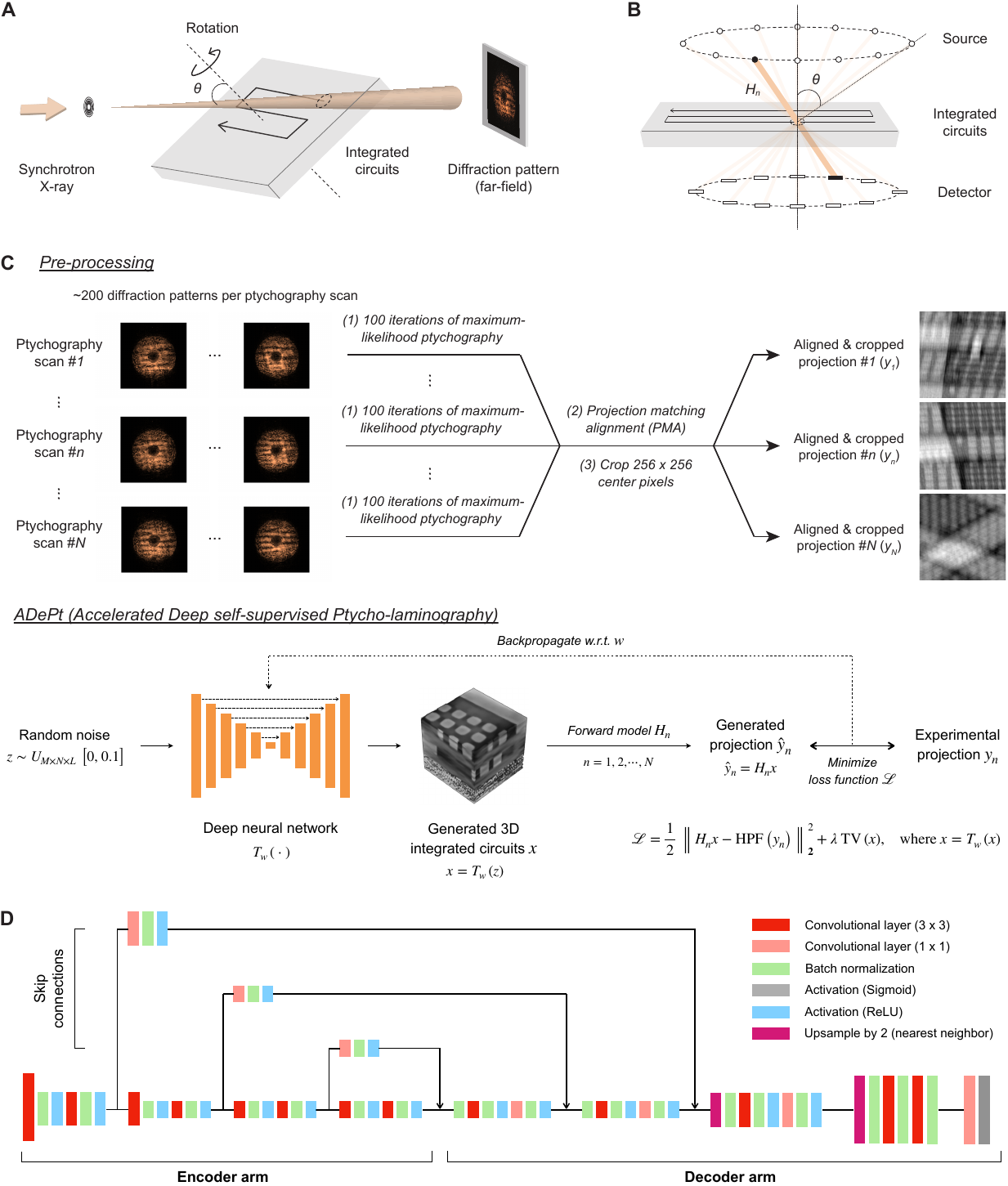}
    \caption{\textbf{Accelerated deep self-supervised ptycho-laminography.} (A) Ptycho-laminographic imaging geometry. Synchrotron X-rays illuminate a sample of integrated circuits in the ptycho-laminography geometry, with the sample rotating around the oblique laminographic axis and scanned over a few thousand angles. For each ptychography scan, the sample is laterally scanned at several hundred different locations. (B) Equivalent imaging geometry. Forward operators $H_n\:(n=1,2,\cdots,N)$ are defined according to each laminographic rotation. (C) Proposed physics-informed machine learning framework. Our pre-processor translates experimental ptycho-laminographic measurement from the detector plane to the sample domain with minimal processing using a ptychographic reconstruction algorithm. ADePt generates a three-dimensional image of integrated circuit morphology from the pre-processed projections throughout the optimization process. (D) Deep neural network architecture for self-supervised learning. The proposed architecture is essentially an encoder-decoder convolutional neural network with skip connections and receiving random noise as input. The output is the image. Code is publicly available at \url{https://github.com/iksungk/ADePt}.}
    \label{fig:introduction}
\end{figure}

As ICs have a flat geometry and an extended field of view, ptychography becomes synergistic with laminography in imaging such nanostructures. Holler \textit{et al} demonstrated ptychographic X-ray laminography on ICs fabricated with $16$\mbox{-}nm technology with $18.9$-nm resolution~\cite{holler2019three} using Laminographic Nano-Imaging instrument (LamNI)~\cite{holler2020lamni}, whose reconstruction quality easily surpassed that of X-ray ptychographic tomography~\cite{holler2017high}. In a typical X-ray ptycho-laminographic imaging apparatus as depicted in Fig.~\ref{fig:introduction}A, synchrotron X-rays obliquely illuminate ICs with the angle of $\theta$ between the direction of X-ray propagation and the rotation axis. ICs are scanned from a few thousand laminographic angular views, where angular sampling depends on the sample thickness, the resolution, and the laminographic angle~\cite{holler2019three}, and for each ptychography scan, X-rays laterally scan the ICs at several-hundred different lateral positions. 

Three-dimensional information of nanostructures is retrieved using a two-step iterative update process based on our ptycho-laminographic measurements, \textit{i.e.} the reconstruction from the densely-sampled dataset (or \textit{densely-sampled reconstruction}): (1) a thousand iterations of a ptychographic reconstruction algorithm processes hundreds of diffraction patterns to get a projection for each of the $2000$ ptychograms (More details about the reconstruction in Sec.~\ref{subsec:reconstruction}.); and (2) a volumetric reconstruction is formed by the laminographic synthesis of all projections. The missing cone in the Fourier domain is filled in during the post-processing step. The densely sampled reconstruction, however, results in a long data acquisition and computation time due to the strict angular sampling requirement~\cite{holler2019three} of ptychograms and the iterative reconstruction of projections.

Here, we demonstrate up to an aggregate $9.57$-fold time savings in X-ray ptycho-laminographic reconstruction by using a physics-regularized deep self-supervised learning architecture, called Accelerated Deep self-supervised Ptycho-laminography (ADePt). We achieved this acceleration by reducing the number of angular samples by $16$ times and the computation time by $4.67$ times. Considering the memory limitations in GPU, ADePt yields the reconstruction of $4.36\times 4.36\times 3.92\mbox{-}\mu\mathrm{m}^3$ ICs within two hours, see Table S1 for more timing details. We provide both quantitative and qualitative comparisons on the performance of ADePt with the densely sampled reconstruction using Bit-Error Ratio (BER) and three-dimensional power spectral density. Finally, we observe that the self-supervised learning kernel fills missing cones from much fewer number of projections compared to the densely sampled reconstruction.

ADePt consists of two components: a lightweight pre-processor and a deep neural network as illustrated in Fig.~\ref{fig:introduction}C. The pre-processor consists of a few iterations of maximum likelihood ptychography, projection matching alignment, and center cropping, and works as an approximate inverse operator on diffraction patterns. The network parameterizes a three-dimensional structure of ICs by considering both network-structure and physics priors. The proposed learning approach makes significant advances from the baseline in terms of the following: (1) ADePt does not use the ground truth structure of ICs for its learning process, which sets it apart from any supervised machine learning approaches and saves significant amount of time and resources from the ground truth preparation, and only operates on a single set of experimental measurements; and (2) ADePt explicitly leverages the physical forward model of the X-ray ptycho-laminography geometry, which tightly regularizes the solution space and guides the algorithm to generate a physically-feasible reconstruction even when the number of projections is largely reduced.

\section{Methods}
\subsection{X-ray ptycho-laminography experiment}
Ptycho-laminography measurements were carried out at the coherent small-angle X-ray scattering (cSAXS) beamline at the Swiss Light Source at the Paul Scherrer Institut (PSI), Switzerland. An integrated circuit produced with $16\mbox{-}\mathrm{nm}$ technology was scanned with synchrotron X-rays of $6.2\:\mathrm{keV}$ using laminographic nano-imaging instrument (LamNI)~\cite{holler2020lamni}. The laminographic angle between the rotation axis and the beam propagation axis was fixed to be $61^\circ$, and the angular step of projections was $0.18^\circ$, making $2000$ scans in total. Diffraction patterns were recorded by a step scan at around $200$ lateral locations with an in-vacuum Eiger 1.5~M detector (pixel size: $75\:\mu\mathrm{m}$, sample-to-detector distance: $5.23\:\mathrm{m}$, exposure time: $0.1\:\mathrm{s}$)~\cite{guizar2014high}.

\subsection{Baseline method and densely sampled reconstruction preparation}\label{subsec:reconstruction}
Densely sampled reconstruction preparation takes experimental ptycho-laminographic measurements from $2000$ ptychography scans and uses a three-step baseline method: (1) For every ptychography scan, projections are retrieved by $1000$ iterations of the least-square likelihood ptychographic algorithm~\cite{odstrvcil2018iterative} as implemented in PtychoShelves~\cite{wakonig2020ptychoshelves}. This ptychographic reconstruction step takes far-field X-ray diffraction patterns ($512\times 512\:\mathrm{px}^2$, pixel size: $75\:\mu\mathrm{m}$) recorded at $200$ different lateral locations. (2) The projections are precisely aligned to each other using the projection matching alignment (PMA) algorithm~\cite{odstrvcil2019alignment}. (3) The aligned projections are synthesized to reconstruct a volumetric structure using the standard Fourier Backprojection (FBP) method, followed by the recovery of missing cone information. Data processing exactly follows the steps in Ref.~\cite{holler2019three}. These steps result in the densely sampled reconstruction with the voxel size of $27.2\:\mathrm{nm}$, which is larger than the resolution of $19\:\mathrm{nm}$ reported in Ref.~\cite{holler2019three} due to fewer number of projections acquired and a larger step size for each scan.

Despite the intensive densely-sampled reconstruction, the densely sampled reconstruction remains ambiguous especially for longitudinal layer features as a missing cone exists in the $k$-space due to the oblique X-ray illumination in the ptycho-laminographic imaging~\cite{holler2019three}.

\subsection{Architecture}
The network design is based on a modified U-net architecture with a random input noise, following the implementation of deep image prior~\cite{ulyanov2018deep}, which has been widely used for many computational imaging applications, including image dehazing~\cite{gandelsman2019double}, super-resolution~\cite{ulyanov2018deep,mataev2019deepred}, phase retrieval~\cite{wang2020phase,bostan2020deep}, tomography~\cite{gong2018pet,baguer2020computed}, and magnetic resonance imaging (MRI)~\cite{liu2020rare}. This implementation shares a similarity with coordinate-based learning approaches~\cite{sitzmann2020implicit,sun2021coil,mildenhall2021nerf,liu2022recovery}. The deep neural network is implemented as an encoder-decoder architecture with skip connections~\cite{ronneberger2015u} as shown in Fig.~\ref{fig:introduction}D. Network weights are initialized with Xavier uniform distribution~\cite{glorot2010understanding} (gain: $0.2$). Following the convention of deep image prior~\cite{ulyanov2018deep}, a random input noise $z$ is given to the network, sampled from a uniform distribution $z\sim U_{M\times N\times L}\left[0,\:0.1\right]$. The encoder reduces the lateral dimensions by a factor of $4$, and the decoder restores the dimensions back to the original. The architecture is chosen to have skip connections to relay the encoder features to the decoder according to our finding that an hourglass architecture does not reliably render high-frequency details. Finally, a sigmoid-like activation function sets the range of output values of the ICs within $\left[-0.03,\:0.03\right]$.

\subsection{Pre-processing and network optimization}
Figure~\ref{fig:introduction}C shows our pre-processor into three steps: (1) $10$-times fewer iterations, \textit{i.e.} $100$, of the least-square maximum likelihood ptychographic algorithm are run to obtain intermediate projections; (2) the projection matching alignment algorithm aligns the intermediate projections to match with each other, which is applied to both densely-sampled and reduced datasets; and (3) $256\times 256$ center pixels are cropped from the aligned projections to be used for the reconstruction process with ADePt. A discussion of the sensitivity to the alignment of projections is summarized in Fig.~S1 in Supplementary Materials.

ADePt iteratively updates the randomly initialized weights in the deep neural network according to the loss functional
\begin{equation}\label{eq:loss_function}
    \hat{w} = \text{argmin}_w\left[\frac{1}{2}\sum_{n=1}^N\left\lVert H_n x - \mathrm{HPF}\left(y_n\right)\right\rVert_\mathbf{2}^2 + \lambda\mathrm{TV}\left(x\right)\right],\:\:\:\mathrm{where}\:\:x=T_w\left(x\right).
\end{equation}
$T_w\left(\cdot\right)$ is a deep neural network parameterized with $w$, $H_n$ a physical forward model corresponding to the $n$-th ptychography scan, $y_n$ a pre-processed projection of the $n$-th ptychography scan, $\mathrm{HPF}\left(\cdot\right)$ a high-pass filter on $y_n$'s, and $\mathrm{TV}\left(\cdot\right)$ a total-variation regularization operator acting upon the $3$D IC structure $x$. The TV regularization parameter $\lambda$ is set differently across the $z$-axis, \textit{i.e.} $\lambda = 3\times 10^{-6}$ for $z < 2.75\:\mu\mathrm{m}$ and $\lambda = 3\times 10^{-8}$ otherwise. Please see Fig.~S2 in Supplementary Material for more details.

The reconstruction process is run for $1500$ iterations using Adam optimizer~\cite{kingma2014adam} with $\beta_1 = 0.9$, $\beta_2 = 0.999$, and the initial learning rate of $2\times 10^{-4}$ which is halved after $1000$ iterations.

\section{Results}
\subsection{Qualitative performance comparison}
\begin{figure}[htbp!]
    \centering
    \includegraphics[width=\textwidth]{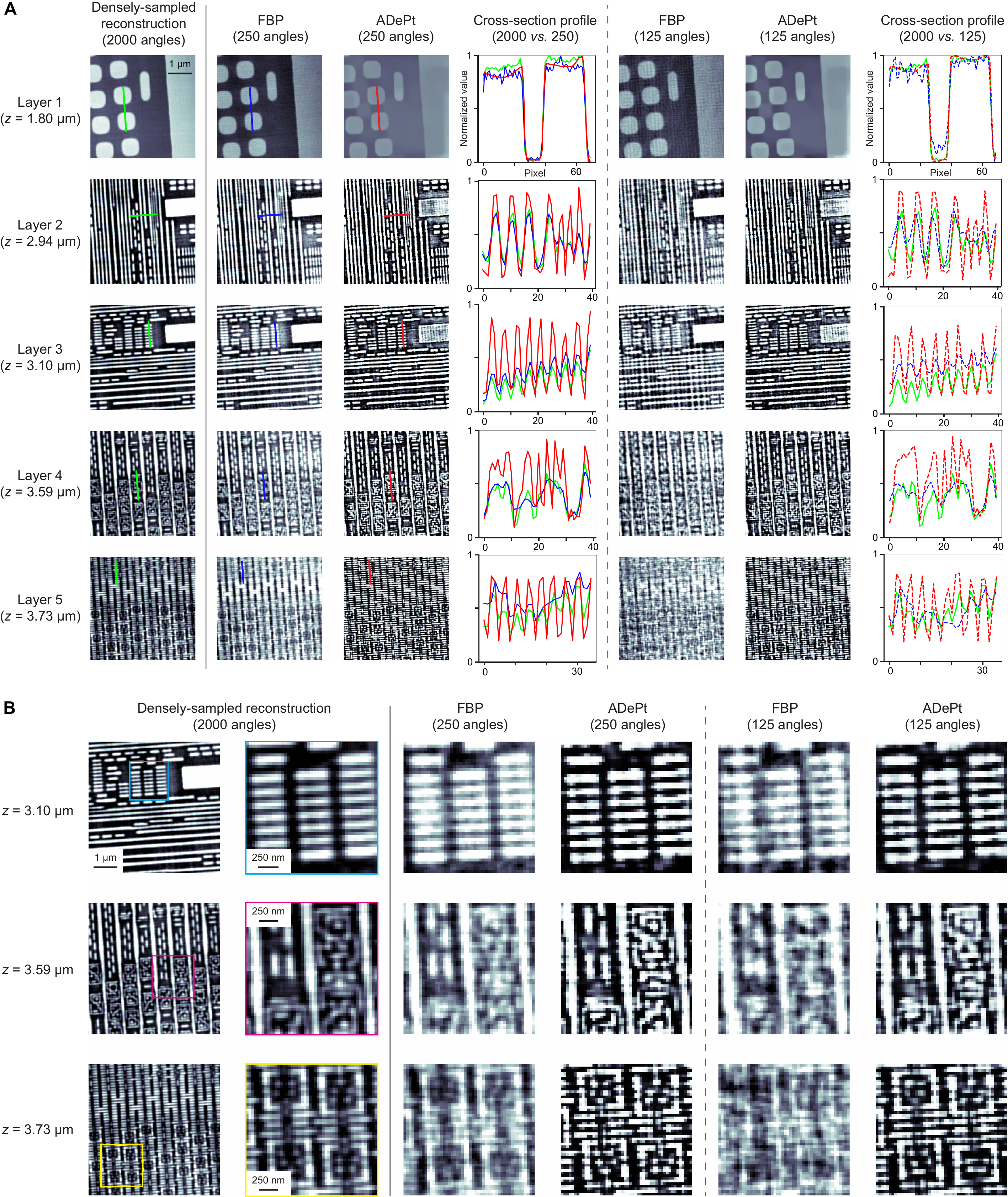}
    \caption{\textbf{Comparison between FBP \& ADePt's reconstructions and the densely sampled reconstruction.} (A) We exploit physics-informed machine learning to reliably reconstruct integrated circuits with a reduced number of projections, \textit{i.e.} $125$ and $250$ out of $2000$, and qualitatively compare the FBP \& ADePt's reconstructions with the densely-sampled reconstruction. Please note that the colormaps used in some reconstructions are selected differently. Specifically, for (1) the densely-sampled reconstruction, the colormaps of Layers 2-5 are set to its 2.5th and 80th percentiles, and for (2) the FBP reconstructions with both 125 and 250 angles, the colormaps of Layers 2-5 are fixed to its 12.5th and 80th percentiles. For all other figures, the colormaps are fixed to the minimum and maximum values of their respective reconstructions. (B) We qualitatively compare the reconstructions within their respective zoomed-in areas for better evaluation. The colormap conventions followed in (A) are also applied.}
    \label{fig:dip_baseline_qualitative_comparison}
\end{figure}

\begin{figure}[htbp!]
    \centering
    \includegraphics[width = \textwidth]{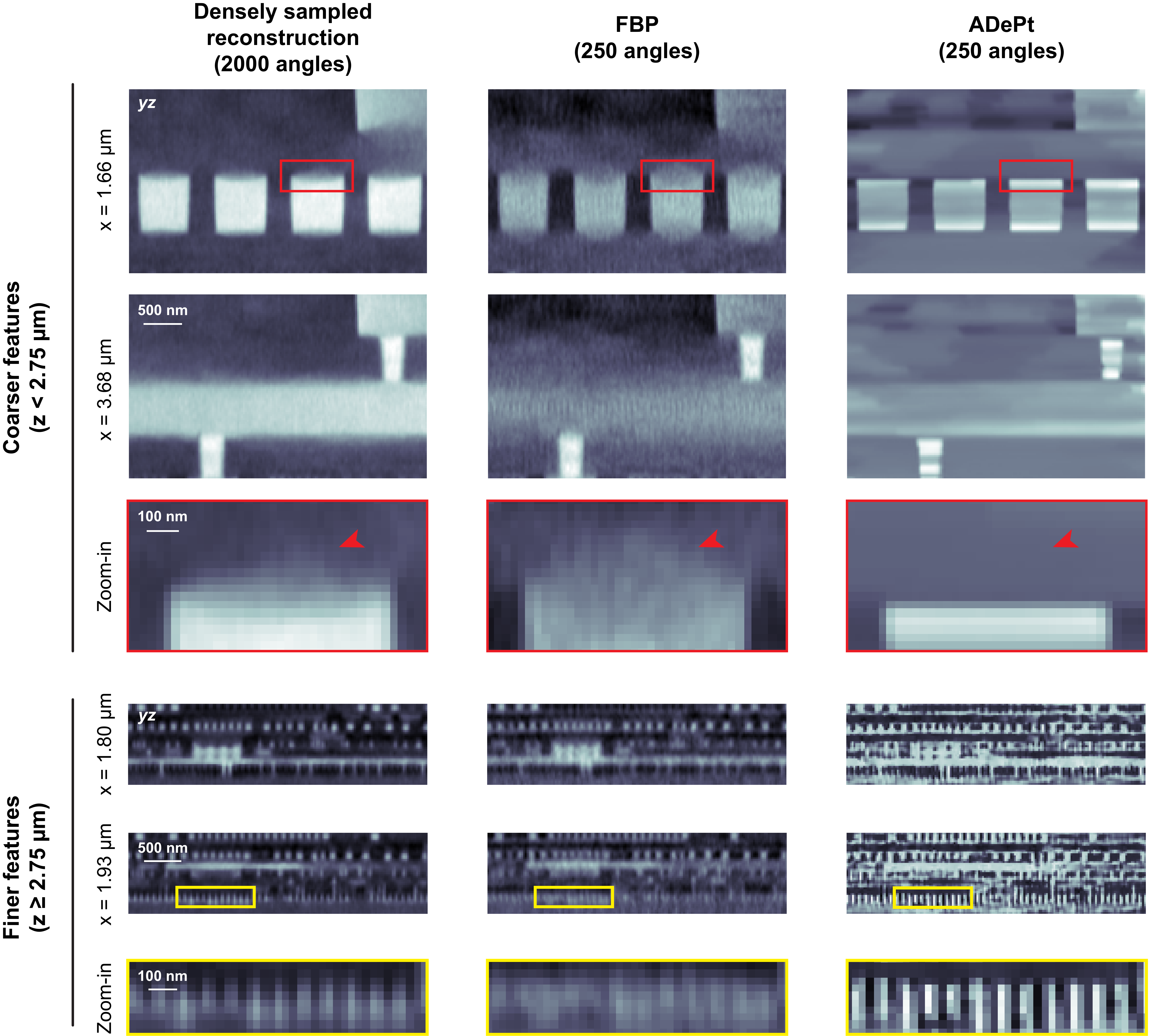}
    \caption{\textbf{Qualitative comparison among $yz$ cross-sections of different reconstructions.} Although the densely sampled reconstruction yields the best contrast among coarser features, missing cone artifacts are not completely addressed (see red arrows). Some finer features can be displayed more clearly with ADePt than with the FBP reconstructions and the densely-sampled reconstruction, at the cost of quantitativeness (see yellow boxes).}
    \label{fig:yz_qualitative_comparison}
\end{figure}

We prepare the densely sampled reconstruction for the performance comparison using the two-step iterative method with $2000$ projections sampled every $0.18^\circ$ angular increment $\Delta\varphi$. Ptycho-laminographic measurements are processed with $1000$ iterations of the least-square maximum likelihood algorithm for each ptychography scan, followed by the projection matching alignment, the filtered backprojection (FBP), and the recovery of missing cone information~\cite{holler2019three}.

In this study, we assess ADePt's three-dimensional physically-feasible rendering performance under sparsely sampled conditions by making a comparison between the densely sampled reconstruction and the ADePt's reconstructions. Here, we increase $\Delta\varphi$ from $0.18^\circ$ to $1.44^\circ$ and $2.88^\circ$, limiting the number of available projections to $250$ and $125$ from $2000$. In Fig.~\ref{fig:dip_baseline_qualitative_comparison}, we visualize reconstructed IC features at five different depths, \textit{i.e.} $z$-axis locations, using the two-step baseline and ADePt with $250$ and $125$ projections to qualitatively compare them with the densely sampled reconstruction. Visually comparing the two, ADePt's reconstructions are more closely aligned to the densely sampled reconstruction for both cases -- even better than the densely sampled reconstruction with greater feature contrast to identify high spatial-frequency details that comes at the cost of quantitativeness -- although the FBP reconstructions show much lower spatial resolution due to accrued artifacts from the missing cones and sparse sampling. We support the argument with the aid of the cross-section profiles. Fig.~S3 in Supplementary Materials visualizes the reconstructions from different axial views. Moreover, ADePt benefits from the sparse sampling scheme, resulting in $\times 6.58$ and $\times 9.57$ aggregate reduction in computation time with $250$ and $125$ projections, respectively, compared to the densely sampled reconstruction. For more details on time breakdown, please see Table~\ref{tab:time_breakdown}.

\begin{table}[htbp!]
    \centering
    \begin{tabular}{cccc}
        \toprule
         & \makecell{Densely sampled\\reconstruction\\(2000 angles)} & \makecell{ADePt\\(250 angles)} & \makecell{ADePt\\(125 angles)}\\
        \midrule
        \textbf{Data acquisition$^\ast$} & $11.1$ & $1.38\:\:(\times 8\:\:\mathrm{faster})$ & $0.694\:\:(\times 16\:\:\mathrm{faster})$\\
        \midrule
        \textbf{Data computation$^\dag$} & $4.27$ & $0.958\:\:(\times 4.46\:\:\mathrm{faster})$ & $0.914\:\:(\times 4.67\:\:\mathrm{faster})$\\
        \footnotesize{\makecell{(1) Iterative update\\(maximum likelihood)}} & \footnotesize{$4.05^\S$} & \footnotesize{$0.0890$} & \footnotesize{$0.0445$}\\
        \footnotesize{\makecell{(2) Reconstruction - \\Data I/O, alignment, synthesis}} & \footnotesize{$0.218^\P$} & \footnotesize{$0.869$} & \footnotesize{$0.869$}\\
        \midrule\midrule
        \textbf{Total time (hr)} & $15.4$ & $2.34\:\:(\times 6.58\:\:\mathrm{faster})$ & $1.61\:\:(\times 9.57\:\:\mathrm{faster})$\\
        \bottomrule
    \end{tabular}
    \caption{\textbf{Data preparation and computation time breakdown.} $^\ast$Data acquisition uses LamNI~\cite{holler2020lamni} to acquire ptycho-laminographic measurements over the integrated circuit sample ($26.2\times 38.2\times 3.92\:\mu\text{m}^3$). $^\dag$Data computation includes an iterative ptychographic update step, data I/O, projection matching alignment~\cite{odstrvcil2019alignment}, and laminographic synthesis. Densely sampled reconstruction uses $1000$ iterations of the LSQ-ML algorithm~\cite{odstrvcil2018iterative,wakonig2020ptychoshelves}, and ADePt's reconstructions use $100$ iterations of the same algorithm. The data computation time breakdown comparison is made on the reconstructed sample ($4.36\times 4.36\times 3.92\:\mu\text{m}^3$). $^\S$Iterative ptychographic update of the densely sampled reconstruction is performed on $10$ GTX 1080 GPUs. $^\P$Data reconstruction process of the densely sampled reconstruction is based on $1$ V$100$ GPU. All other computations use $2$ V$100$ GPUs.}
    \label{tab:time_breakdown}
\end{table}

\subsection{Power spectral density representation}
It is easier to describe the aforementioned artifacts in a different domain as both the terms sparse sampling and missing cones are defined based on the $k$-space. We visualize both the densely sampled reconstruction and FBP \& ADePt reconstructions as $k$-space representations by means of power spectral density, which is the Fourier transform of the autocorrelation function, \textit{i.e.} the Wiener-Khinchin theorem.

\begin{figure}[htbp!]
    \centering
    \includegraphics[width=\textwidth]{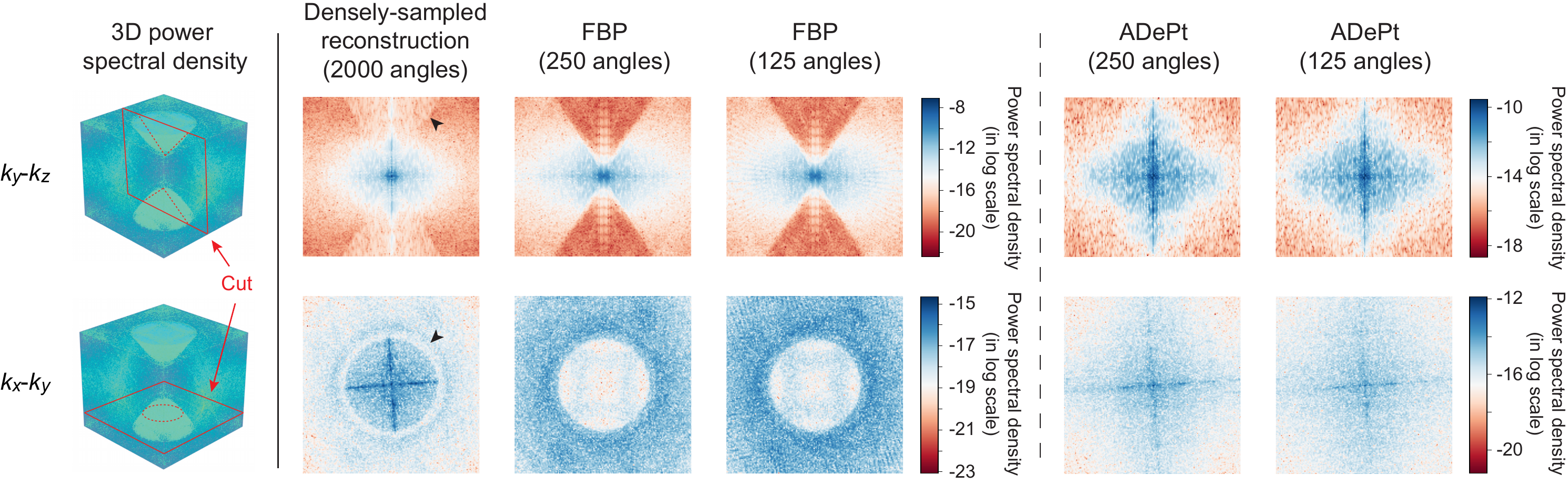}
    \caption{\textbf{Power spectral density analysis for qualitative comparison.} We visualize FBP \& ADePt's reconstructions and the densely sampled reconstruction in $k$-space to visualize artifacts due to missing cone and sparse sampling. Cuts are made along the $k_y\mbox{-}k_z$ and $k_x\mbox{-}k_y$ planes (red arrows). We demonstrate that ADePt provides reconstructions with fewer artifacts in $k$-space, considering that the FBP reconstructions display artifacts due to the angular subsampling and missing cone, and that the densely-sampled reconstruction shows artifacts due to imperfect missing cone filling (black arrows).}
    \label{fig:psd_comparison}
\end{figure}

In Fig.~\ref{fig:psd_comparison}, we provide two-dimensional power spectral density profiles of the baseline and ADePt's reconstructions using $250$ and $125$ projections along with the densely sampled reconstruction. Both FBP reconstructions show their missing cones rooted in an oblique rotation axis in the ptycho-laminography geometry. Sparse-sampling artifacts are also easily noticed in the reconstructions, which provides explanations on lack of spatial resolution as shown in Fig.~\ref{fig:dip_baseline_qualitative_comparison}.

Our framework, however, learns from experimental ptycho-laminographic measurements how the missing parts should be filled to represent physically feasible IC solutions guided by the physical and network-structure priors. Moreover, comparing the densely sampled reconstruction and ADePt's reconstructions in $k$-space, we notice that ADePt renders ICs even better -- the missing cone of the densely sampled reconstruction remain unintentionally accentuated, as shown with black arrows in Fig.~\ref{fig:psd_comparison}, whereas the cone filled and spectrum recovered with ADePt is more continuous. As already reported elsewhere~\cite{zhou2020diffraction}, it is plausible that ADePt could outperform the baseline method even with more limited amount of data given as the proposed framework takes a self-supervised approach, thus completely agnostic to the type of specimen. However, this does not preclude future use of priors to boost performance.

\subsection{Quantitative evaluation}\label{sec:quantitative_analysis}
\begin{figure}[htbp!]
    \centering
    \includegraphics[width=\textwidth]{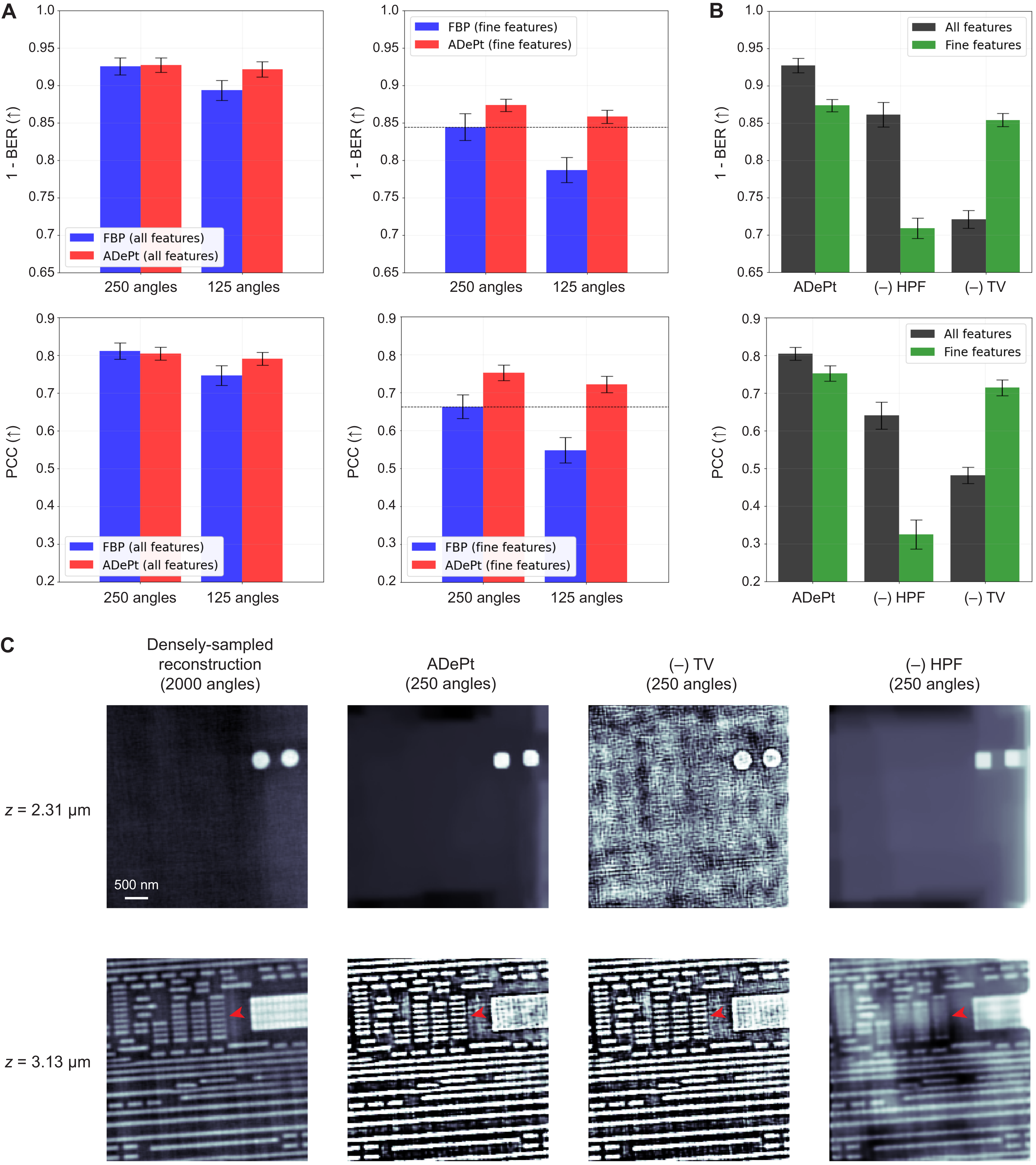}
    \caption{\textbf{Quantitative analysis of reconstructions and ablation study.} (A) We use bit-error rate (BER) and Pearson correlation coefficient (PCC) to compare ADePt's reconstructions with the baseline for different scales of features in the integrated circuits. (B) Ablation study. We assess relative contribution of each design element, \textit{i.e.} high-pass filtering (HPF) and total-variation (TV) regularization, to the final reconstruction by removing one at a time from the complete model. We incorporate HPF to enforce a high-frequency content bias to our deep neural network and TV regularization to suppress spurious high-frequency artifacts in the background. Fig.~S4 in Supplementary Materials illustrates (A) and (B) further using another quantitative metric. (C) The self-supervised learning algorithm behaves unfavorably when each component is ablated. Total variation (TV) regularization suppresses high-frequency artifacts, and high-pass filtering (HPF) improves the spatial resolution of features recovered by the algorithm (red arrows).}
    \label{fig:quantitative_ablation_architecture}
\end{figure}

We use Bit-Error Rate (BER) as a metric to quantify the ratio of erroneous occupancy in the reconstructions with reference to the densely sampled reconstruction as the ground truth~\cite{kang2022attentional}. Binary objects such as integrated circuits are particularly suited to this metric for quality assessment. Although many printing materials comprise ICs, including copper, tungsten, and aluminum, here we just treat the ICs as binary with regard to the occupancy, irrespective of material. Using the Expectation-Maximization (EM) algorithm in the context of Gaussian mixture models, we first binarize the densely-sampled reconstruction to get the ground truth. However, since the densely-sampled reconstruction may still be ambiguous especially for longitudinal features due to missing cone in the Fourier domain (see Fig.~\ref{fig:psd_comparison}), the layers with ambiguous features will not be accurately binarized. Thus, we exclude the layers with binarization errors from our quantitative analyses in Section~\ref{sec:quantitative_analysis}. More details can be found in Fig.~S5 in Supplementary Materials.

In Fig.~\ref{fig:quantitative_ablation_architecture}A, we compare ADePt's performance with the baseline method's using BER and Pearson correlation coefficient (PCC). We demonstrate that ADePt outperforms the baseline method for both conditions with fewer projections, but the difference between two methods becomes more statistically significant when only $125$ projections are considered during the reconstruction process, which suggests that ADePt is more tolerant to sparse sampling. The layers located at $z > 2.75\:\mu\mathrm{m}$ are considered to have finer circuit features, according to typical design rules of integrated circuits (Fig.~S2 in Supplementary Materials illustrates this further). This implies that higher-frequency details are better reconstructed with the physics-informed self-supervised machine learning, even at regions with fine transverse details.

\subsection{Ablation study}
As illustrated in the loss function in Fig.~\ref{fig:introduction}C, we employ two key design elements to the implementation: (1) high-pass filtering on the intermediate projections to improve feature contrast and to enforce high-frequency bias to the network; and (2) total-variation regularization in the loss functional to suppress residual artifacts. We assess the relative contribution of each design element in our proposed framework by ablating each one of the following elements in succession: high-pass filter on projections and total-variation regularization. 

The results are shown in detail in Fig.~\ref{fig:quantitative_ablation_architecture}. One salient observation is that in the region of all features ($z$ between $0$ and $3.92~\mu\text{m}$) performance degrades by equal amounts if the high-pass filter or TV term are ablated. On the other hand, in the fine feature regime ($z$ between $2.75$ and $3.92~\mu\text{m}$) the significance of the high-pass filter is higher, as evidenced in Fig.~\ref{fig:quantitative_ablation_architecture}B by the catastrophic drop in performance when it is ablated. Visual inspection of the ablated reconstructions in Fig.~\ref{fig:quantitative_ablation_architecture}C confirms these trends.

\section{Discussion}
ADePt provides a three-dimensional estimate of integrated circuits from experimental ptycho-laminographic measurements using deep self-supervised learning. The proposed framework makes explicit use of a physical forward model to obtain the image iteratively, and regularizes through the chosen sparsity-enforcing neural network kernel and an additional total variation penalty in the loss function. This is a significant departure from supervised learning approaches that typically require ground truth to prepare a paired dataset for training, which is expensive. 

Supervised methods' performance is also limited in the absence of accurate ground truth. For IC imaging this is almost always the case since obtaining the true shape is challenging. In earlier supervised work~\cite{kang2022attentional}, we used reconstructions from dense angular sampling as ground truth. Here, we observe that ADePt fills the missing cone more effectively than the densely sampled reconstructions, as Fig.~\ref{fig:psd_comparison} clearly illustrated. That the integrated circuit geometry is very compatible with our chosen auxiliary total variation regularizer strengthens this claim. Further investigation of the ADePt scheme's performance in different types of specimens is a good topic for future work. 

In this work, we treat all ptycho-laminographic projections as equal. However, in previous work~\cite{kang2021dynamical} we have demonstrated that there is benefit to weighing each differently, according to an attentional scheme~\cite{vaswani2017attention,wang2020axial}. Also, coordinate-based learning methods~\cite{sun2021coil,sitzmann2020implicit,mildenhall2021nerf,liu2022recovery} may be beneficial to further increase the reconstruction volume as the methods generally take fewer trainable parameters than convolutional neural network architectures. We leave this for future work as well.

\section*{Backmatter}

\begin{backmatter}
\bmsection{Funding}
U.S. Department of Energy (DE-AC02-06CH11357); Korea Foundation for Advanced Studies; Intelligence Advanced Research Projects Activity (FA8650-17-C-9113).

\section*{Acknowledgement}
We are grateful to William Harrod, Ed Cole, Lee Oesterling, Antonio Orozco, and Yudong Yao for helpful discussions and comments, and acknowledge Los Alamos National Laboratory (LANL)'s contributions. Funding from the Intelligence Advanced Research Projects Activity, Office of the Director of National Intelligence (IARPA-ODNI), contract FA8650-17-C-9113 is gratefully acknowledged. The MIT SuperCloud and Lincoln Laboratory Supercomputing Center provided resources (high performance computing, database, consultation) that have contributed to the research results reported within this paper. I. Kang also acknowledges support from Korea Foundation for Advanced Studies (KFAS). This research used resources of the Advanced Photon Source, a U.S. Department of Energy (DOE) Office of Science User Facility, operated for the DOE Office of Science by Argonne National Laboratory under Contract No. DE-AC02-06CH11357. The views and conclusions contained herein are those of the authors and should not be interpreted as necessarily representing the official policies or endorsements, either expressed or implied, of the ODNI, IARPA or the US Government. The measurements were performed at the cSAXS beamline of the Swiss Light Source at the Paul Scherrer Institut, Switzerland. Samples were prepared at the University of Southern California.

\bmsection{Disclosures}
The authors declare no competing interests.

\bmsection{Data Availability Statement}
Codes are publicly available at \url{https://github.com/iksungk/ADePt}.

\end{backmatter}

\bibliography{optica_bib}
\end{document}